\documentclass[conference, a4paper]{IEEEtran}
\IEEEoverridecommandlockouts
\def\BibTeX{{\rm B\kern-.05em{\sc i\kern-.025em b}\kern-.08em
    T\kern-.1667em\lower.7ex\hbox{E}\kern-.125emX}}

\usepackage{amsmath,amssymb,amsfonts,graphicx,mathtools,nccmath,amsthm,float}
\usepackage{multicol,tikz,cite,array,booktabs,url}
\usepackage{cite}
\usepackage{textcomp}
\usepackage{xcolor}
\usepackage{algorithm}
\usepackage{algpseudocode,algorithmicx}
\usepackage[top=.75in, left= 0.625in, right = 0.63in, bottom =1in]{geometry}

\algnewcommand{\LineComment}[1]{\State \(\#\) #1}
    
\graphicspath{{figs/}}
\DeclareGraphicsExtensions{.pdf}

\newcommand{\vect}[1]{\boldsymbol{\mathbf{#1}}}

\newtheorem{Cor}{Corollary}
\newtheorem{Lem}{Lemma}

\DeclareMathOperator{\diag}{diag}

\DeclareMathOperator{\trace}{Tr}
\DeclarePairedDelimiter{\norm}{\lVert}{\rVert}

\algnewcommand\algorithmicinput{\textbf{Set}}
\algnewcommand\Set{\item[\algorithmicinput]}
\algnewcommand\algorithmicinitial{\textbf{Initialize}}
\algnewcommand\Initialize{\item[\algorithmicinitial]}

\let\oldReturn\Return
\renewcommand{\Return}{\State\oldReturn}

\begin{document}

\title{Safeguarding MIMO Communications with Reconfigurable Metasurfaces and Artificial Noise
}
\author{\IEEEauthorblockN{George~C.~Alexandropoulos$^1$, Konstantinos~Katsanos$^1$, Miaowen~Wen$^2$, and Daniel B. da Costa$^3$}
$^1$Department of Informatics and Telecommunications, National and Kapodistrian University of Athens, Greece\\
$^2$School of Electronic and Information Engineering, South China University of Technology, China\\
$^3$Department of Computer Engineering, Federal University of Cear\'{a}, Brazil\\
E-mails: \{alexandg, kkatsan\}@di.uoa.gr, eemwwen@scut.edu.cn, danielbcosta@ieee.org}

\maketitle

\begin{abstract}
Wireless communications empowered by Reconfigurable Intelligent (meta)Surfaces (RISs) are recently gaining remarkable research attention due to the increased system design flexibility offered by RISs for diverse functionalities. In this paper, we consider a Multiple Input Multiple Output (MIMO) physical layer security system with multiple data streams including one legitimate and one eavesdropping passive RISs, with the former being transparent to the eavesdropper and the latter's presence being unknown at the legitimate link. We first focus on the eavesdropping subsystem and present a joint design framework for the eavesdropper's combining vector and the reflection coefficients of the eavesdropping RIS. Then, focusing on the secrecy rate maximization, we propose a physical layer security scheme that jointly designs the legitimate precoding vector and the Artificial Noise (AN) covariance matrix, as well as the legitimate combining vector and the reflection coefficients of the legitimate RIS. Our simulation results reveal that, in the absence of a legitimate RIS, transceiver spatial filtering and AN are incapable of offering nonzero secrecy rates, even for eavesdropping RISs with small numbers of elements. However, when a $L$-element legitimate RIS is deployed, confidential communication can be safeguarded against cases with even more than a $5L$-element eavesdropping RIS. 
\end{abstract}


\section{Introduction} \label{Sec:Intro}
Reconfigurable Intelligent (meta)Surfaces (RISs) have been recently envisioned as a revolutionary means to transform any passive wireless communication environment to an active reconfigurable one \cite{Liaskos_Visionary_2018_all, George_RIS_TWC2019_all, Wu_RIS_TWC2019}, offering increased environmental intelligence for diverse communication objectives. A RIS is an artificial planar structure with integrated electronic circuits \cite{Kaina_metasurfaces_2014_all} that can be programmed to manipulate an incoming electromagnetic field in a wide variety of functionalities \cite{Marco_Visionary_2019_all1, HMIMO_all}. Among the various RIS-enabled objectives belongs the Physical Layer Security (PLS) \cite{Yang_ComMag_2015_all}, which is considered as a companion technology to conventional cryptography, targeting at significantly enhancing the quality of secure communication in beyond $5$-th generation (5G) wireless networks.

One of the very first recent studies on RIS-enabled PLS systems is \cite{Chen_2019_all}, which considered a legitimate Multiple Input Single Output (MISO) broadcast system, multiple eavesdroppers, and one RIS for various configurations for the reflection coefficients of its discrete unit elements. In that work, aiming at safeguarding legitimate communication, an Alternating Optimization (AO) approach for designing the RIS phase matrix and the legitimate precoder was presented together with a suboptimal scheme based on Zero Forcing (ZF) precoding that nulls information leakage to the eavesdroppers. In \cite{Shen_2019_all}, the secrecy rate maximization problem was investigated for a RIS-empowered legitimate system comprising a multi-antenna transmitter and a single-antenna receiver in the vicinity of an eavesdropper with multiple antenna elements. Efficient resource allocation algorithms for the case of multiple legitimate receivers and one eavesdropper were presented in \cite{Cui_2019_all, Xu_2019_all, Yu_2019_all, Almohamad_2020}. The MISO secrecy channel with the help of a single legitimate RIS was also considered in \cite{Chu_2020_all}, with the goal to minimize the transmit power subject to a constraint which keeps the secrecy rate above a target value. It was shown by means of computer simulations that RIS deployment leads to transmit power reservation. On the other hand, a new type of attack, termed as RIS jamming attack, was investigated in \cite{Lyu_2020_all}, according to which a passive RIS reflects jamming signals harming legitimate communication. The presented experimental results exhibited that the legitimate received signal can be downgraded up to $99\%$, witnessing that a RIS can be effectively used by the eavesdropping side for zero-power jamming. Very recently in \cite{Hong_2020}, a RIS-assisted Multiple Input Multiple Output (MIMO) PLS system was considered, where the precoding matrix for fixed number of data streams, the Artificial Noise (AN), and the RIS reflection configuration of the legitimate side were jointly designed targeting the secrecy rate maximization.

All above recent studies indicate that RIS-empowered PLS systems are able to offer increased flexibility for both the legitimate and eavesdropping sides, enabling increased secrecy or (cooperative) jamming \cite{Cumanan_2017_all} in efficient ways. In this paper, we study multi-stream MIMO PLS systems with both legitimate and eavesdropping passive RISs. Focusing first on the eavesdropping subsystem, we present a joint design framework for the eavesdropper's combining vector and the reflection coefficients of the eavesdropping RIS. Then, by formulating and solving a novel joint design problem for the legitimate subsystem, we propose a PLS scheme incorporating legitimate precoding and AN, receive combining, and passive beamforming from the legitimate RIS. Differently from \cite{Hong_2020}, the presented optimization framework includes the number of data streams and the legitimate receive combiner. Our simulation results demonstrate that the proposed design can secure confidential communication over eavesdropping RISs with large numbers of elements, outperforming the state-of-the-art techniques in terms of the secrecy rate performance.

\textit{Notations:} Vectors and matrices are denoted by boldface lowercase and boldface capital letters, respectively. The transpose, conjugate, Hermitian transpose and inverse of $\mathbf{A}$ are denoted by $\mathbf{A}^T$, $\mathbf{A}^*$, $\mathbf{A}^H$, and $\mathbf{A}^{-1}$ respectively, and $|\mathbf{A}|$ is the determinant of $\mathbf{A}$, while $\mathbf{I}_{n}$ and $\mathbf{0}_{n}$ ($n\geq2$) are the $n\times n$ identity and zeros' matrices, respectively. ${\rm Tr}(\mathbf{A})$ and $\norm{\mathbf{A}}_F$ represent $\mathbf{A}$'s trace and Frobenius norm, respectively, while notation $\mathbf{A}\succ\vect{0}$ ($\mathbf{A}\succeq\vect{0}$) means that the square matrix $\mathbf{A}$ is Hermitian positive definite (semi-definite). $[\mathbf{A}]_{i,j}$ is the $(i,j)$-th element of $\mathbf{A}$, $[\mathbf{a}]_i$ is $\mathbf{a}$'s $i$-th element of $\mathbf{a}$, ${\rm diag}\{\mathbf{a}\}$ denotes a square diagonal matrix with $\mathbf{a}$'s elements in its main diagonal. $\odot$ and $\otimes$ stand for the Hadamard and Kronecker products, respectively, while $\operatorname{vec}(\vect{A})$ indicates the vector which is comprised by stacking the columns of a matrix $\vect{A}$, and $\operatorname{unit}(\mathbf{a})$ means $\mathbf{a}$ has its elements normalized. $\nabla_{\mathbf{a}}^{\rm R}f$ represents the Riemannian gradient vector of a scalar function $f$ along the direction indicated by $\mathbf{a}$. $\mathbb{C}$ represents the complex number set, $|a|$ denotes the amplitude of the complex scalar $a$, and $\Re(a)$ its real part. $\mathbb{E}\{\cdot\}$ is the expectation operator and $\mathbf{x}\sim\mathcal{CN}(\mathbf{a},\mathbf{A})$ indicates a complex Gaussian random vector with mean $\mathbf{a}$ and covariance matrix $\mathbf{A}$. 

\section{System and Signal Models} \label{Sec:Sys_Model}
\begin{figure}[t!]
\centering
\includegraphics[scale=0.55]{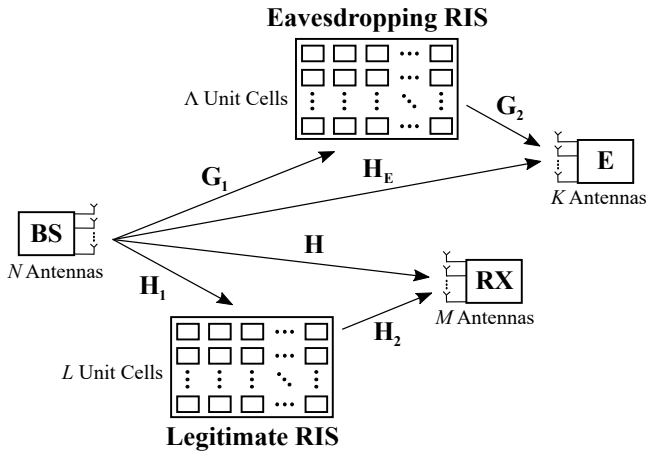}
\caption{The considered PLS system comprising three multi-antenna nodes and two multi-element RISs, one serving the eavesdropper E and the other the legitimate BS-RX link. BS is assumed unaware of the existence of the eavesdropping RIS, the same is assumed for E regarding the legitimate RIS.}
\label{fig:System_Model}
\end{figure}

The considered system model, as illustrated in Fig$.$~\ref{fig:System_Model}, consists of a Base Station (BS) equipped with $N$ antenna elements wishing to communicate in the downlink direction with a legitimate Receiver (RX) having $M$ antennas. This downlink transmission is assumed to be further empowered by a legitimate RIS with $L$ unit cells, which is placed close to RX. In the vicinity of the legitimate BS-RX link exists a $K$-antenna ($K\geq M$) Eavesdropper (E) with an eavesdropping RIS of $\Lambda$ unit elements close to it, that is intended for enabling legitimate information decoding at E's side. We assume that the legitimate RIS is connected to the legitimate node via dedicated hardware and control signaling for online reconfigurability; the same holds for E and the eavesdropping RIS. The BS knows about the existence of E and focuses on securing its confidential link with RX; however, it is unaware of the presence of the eavesdropping RIS. It is also assumed that the deployment of the legitimate RIS is transparent to E. 

We assume throughout this paper that perfect channel information is available at the BS and E sides via pilot-assisted channel estimation. Specifically, BS possesses the channels $\vect{H}\in\mathbb{C}^{M \times N}$, $\vect{H}_1\in\mathbb{C}^{L \times N}$, and $\vect{H}_2\in\mathbb{C}^{M \times L}$ referring to the BS-RX, BS to legitimate RIS, and legitimate RIS links to RX, respectively. It is also assumed that BS and E cooperate in order to both estimate the BS-E channel $\vect{H}_{\rm E}\in\mathbb{C}^{K \times N}$ as follows: BS transmits pilot signals to E that estimates $\vect{H}_{\rm E}$ and then feeds this estimation back to BS. This cooperation may apply to the case where E plays the dual role of a legitimate receiver and of an eavesdropper. Recall that BS is unaware of the existence of the eavesdropping RIS, hence, it has no knowledge on the BS to the eavesdropping RIS channel $\vect{G}_1\in\mathbb{C}^{\Lambda \times N}$ and the eavesdropping RIS to E channel $\vect{G}_2\in\mathbb{C}^{K \times \Lambda}$. However, the latter two channels are assumed available at the E side. It is noted that channels where a RIS is involved can be estimated either at the RIS side with a single active hardware element \cite{George_RIS_2020} or via cascaded channel estimation with pilot reflection patterns at RIS \cite{Deepak_ICASSP_2019}. In the case of an eavesdropping RIS as in \cite{George_RIS_2020}, the BS pilots can be also available to the RIS side, via E that possesses them, in order to enable channel estimation. We finally assume that due to obstacles there are no actual channels between the legitimate RIS and E, and the eavesdropping RIS and RX.  

\subsection{Received Signal Models and Secrecy Rate}
To secure the confidentiality of the legitimate link, BS applies AN \cite{Liu_2017_all} that is jointly designed with the BS precoding scheme, the RX combiner, and the legitimate RIS reflection (passive beamforming) vector $\vect{\phi}\triangleq[e^{j \theta_1}\,\,e^{j \theta_2}\,\,\cdots\,\,e^{j \theta_L} ]^T\in\mathbb{C}^{L \times 1}$, where $\theta_{\ell}$ with $\ell = 1,2,\ldots,L$ denotes the phase shifting value at the $\ell$-th RIS unit element. We represent by $\vect{x} \in \mathbb{C}^{N \times 1}$ the transmitted signal from the BS antenna elements, which is composed as $\vect{x}\triangleq\vect{V} \vect{s} + \vect{z}$, where $\vect{V} \in \mathbb{C}^{N \times N_d}$ is the linear precoding matrix and $\vect{s} \sim \mathcal{CN}(\mathbf{0}_{N_d},\mathbf{I}_{N_d})$ is the legitimate information symbol vector comprised of $N_d \leq \min\{M,N\}$ independent streams, which is assumed independent from the AN vector $\vect{z}\in\mathbb{C}^{N \times 1}$ having the covariance matrix $\vect{Z}\triangleq\mathbb{E}\{\vect{z}\vect{z}^H\}$. The baseband received signal vectors $\vect{y}_{\rm RX}\in\mathbb{C}^{M \times 1}$ and $\vect{y}_{\rm E}\in\mathbb{C}^{K\times 1}$ at the RX and E antennas can be mathematically expressed as
\begin{align}
    \vect{y}_{\rm RX} &= \left(\vect{H} + \vect{H}_2 \vect{\Phi} \vect{H}_1\right) \left( \vect{V} \vect{s} +  \vect{z} \right) + \vect{n}_{\rm RX}, \label{eq:y_RX} \\
    \vect{y}_{\rm E} &= \left(\vect{H}_{\rm E} + \vect{G}_2 \vect{\Psi} \vect{G}_1 \right) \left( \vect{V} \vect{s} +  \vect{z} \right) + \vect{n}_{\rm E}, \label{eq:y_E}
\end{align}
where $\vect{\Phi}\triangleq\diag\{\vect{\phi}\}\in\mathbb{C}^{L \times L}$ and $\vect{\Psi}\triangleq\diag\{\vect{\psi}\}\in\mathbb{C}^{\Lambda \times \Lambda}$ with $\vect{\psi}\triangleq[e^{j \xi_1}\,\,e^{j \xi_2}\,\,\cdots\,\,e^{j \xi_\Lambda} ]^T\in\mathbb{C}^{\Lambda\times 1}$ being the eavesdropping RIS reflection vector in which $\xi_k$ with $k = 1,2,\ldots,\Lambda$ represents the phase shifting value at the $k$-th RIS unit element. In the latter two expressions, $\vect{n}_{\rm RX}\sim \mathcal{CN}(\vect{0}_{M},\sigma^2 \vect{I}_M)$ and $\vect{n}_{\rm E}\sim \mathcal{CN}(\vect{0}_{K},\sigma^2 \vect{I}_K)$ stand for the Additive White Gaussian Noise (AWGN) vectors.

We assume the linear combining matrices $\vect{U} \in \mathbb{C}^{M \times N_d}$ and $\vect{W} \in \mathbb{C}^{K \times N_d}$ for RX and E, respectively, which will be designed later on. This assumption considers that E knows the value of $N_d$ used in the legitimate link, which will serve as an upper bound for E's achievable rate. The achievable rates at the legitimate and eavesdropping links are given by
\begin{equation} \label{eq:R_RX}
\begin{aligned}
 \mathcal{R}_{\rm RX} \triangleq & \log_2 \Bigl| \vect{I}_{N_d} \!+\! \vect{U}^H \tilde{\vect{H}} \vect{V}\vect{V}^H \tilde{\vect{H}}^H \vect{U} \\
 &\times \left( \vect{U}^H \left( \sigma^2 \vect{I}_M \!+\! \tilde{\vect{H}} \vect{Z} \tilde{\vect{H}}^H  \right) \vect{U} \right)^{-1} \Bigr|,
\end{aligned}    
\end{equation}
\begin{equation} \label{eq:R_E}
\begin{aligned}
 \mathcal{R}_{\rm E} \triangleq & \log_2 \Bigl| \vect{I}_{N_d} \!+\! \vect{W}^H \tilde{\vect{H}}_{\rm E} \vect{V}\vect{V}^H \tilde{\vect{H}}_{\rm E}^H \vect{W} \\ &\times \left( \vect{W}^H \left( \sigma^2 \vect{I}_K \!+\! \tilde{\vect{H}}_{\rm E} \vect{Z} \tilde{\vect{H}}_{\rm E}^H  \right) \vect{W} \right)^{-1} \Bigr|,
\end{aligned}
\end{equation}
where $\tilde{\vect{H}}\triangleq\vect{H} + \vect{H}_2 \vect{\Phi} \vect{H}_1$ and $\tilde{\vect{H}}_{\rm E}\triangleq\vect{H}_{\rm E} + \vect{G}_2 \vect{\Psi} \vect{G}_1$. The secrecy rate is then obtained as $\mathcal{R}_{\rm s}\triangleq\max\{0,\mathcal{R}_{\rm RX}-\mathcal{R}_{\rm E}\}$.

\subsection{Design of the Eavesdropping Parameters $\vect{W}$ and $\vect{\psi}$}\label{E_design}
We assume that E is unaware of the fact that BS transmits the AN vector $\vect{z}$, and jointly designs $\vect{W}$ and $\vect{\psi}$ profiting from the availability of the channels $\vect{H}_{\rm E}$, $\vect{G}_1$, and $\vect{G}_2$. To this end, E considers that BS performs ZF precoding to null $\vect{H}_{\rm E}$, as such, it assumes that its baseband received signal is given by $\bar{\vect{y}}_{\rm E}\triangleq \bar{\vect{H}}_{\rm E} \vect{V} \vect{s} + \vect{n}_{\rm E}$ (and not the correct expression \eqref{eq:y_E} including AN), where $\bar{\vect{H}}_{\rm E}\triangleq\vect{G}_2 \vect{\Psi} \vect{G}_1$, while for the considered $\vect{V}$'s columns holds $N_d\leq N-K$. It then formulates the following joint design optimization problem:
\begin{align*}
    \mathcal{OP}_{\rm E}: \,\max_{{\vect{W}, \, \vect{\psi}}} \,\bar{\mathcal{R}}_{\rm E},  \text{s.t.} \,\, \norm{\vect{W}}_F^2 \leq 1,\, \lvert \psi_{k} \rvert = 1 \, \forall k = 1,2,\dots,\!\Lambda,
\end{align*}
where the rate $\bar{\mathcal{R}}_{\rm E}$ is given using $\mathbf{C}\triangleq\vect{W}^H \vect{W}$ by:
\begin{equation}
\bar{\mathcal{R}}_{\rm E} \triangleq \log_2 \left| \vect{I}_{N_d} + \sigma^{-2} \vect{W}^H\bar{\vect{H}}_{\rm E}  \vect{V} \vect{V}^H \bar{\vect{H}}_{\rm E}^H \vect{W} \mathbf{C}^{-1} \right|.
\end{equation}
To solve this problem we adopt AO similar to the approach that will be described in the following section. The detailed solution of $\mathcal{OP}_{\rm E}$ will be presented in an extended version of this paper.

\section{Proposed RIS-Empowered Secrecy Design} \label{Sec:Prob_Form}
According to the considered system model, BS lacks knowledge about the existence of any eavesdropping RIS. Hence, its believed baseband received signal at E given the availability of $\vect{H}_{\rm E} $ at its side is $\hat{\vect{y}}_{\rm E} \triangleq \vect{H}_{\rm E} \left( \vect{V} \vect{s} + \vect{z} \right) + \vect{n}_{\rm E}$, instead of the actual signal model in \eqref{eq:y_E}. Using the latter expression and assuming capacity-achieving combining at E, the BS formulates E's achievable rate as the following function of $\vect{V}$ and $\vect{Z}$:
\begin{equation} \label{eq:R_E_bs}
 \hat{\mathcal{R}}_{\rm E} \triangleq \log_2 \left\lvert \vect{I}_K \!+\! \vect{H}_{\rm E} \vect{V} \vect{V}^H \vect{H}_{\rm E}^H \left( \sigma^2 \vect{I}_K \!+\! \vect{H}_{\rm E} \vect{Z} \vect{H}_{\rm E}^H \right)^{-1} \right\rvert.
\end{equation}

In this paper, we consider the following secrecy rate maximization problem for the joint design of the legitimate BS linear precoding vector $\vect{V}$ and the number of streams $N_d$, the AN covariance matrix $\vect{Z}$, the linear combiner $\vect{U}$ at RX, and the reflection vector $\vect{\phi}$ of the legitimate RIS:
\begin{align*}
\begin{split}
    \mathcal{OP}_{\rm L}: \,\, &\max_{N_d, \vect{U}, \vect{V}, \vect{Z} \succeq \vect{0},  \vect{\phi}} \quad \hat{\mathcal{R}}_{\rm s} \triangleq \mathcal{R}_{\rm RX} - \hat{\mathcal{R}}_{\rm E} \\
    &\text{s.t.} \quad \trace(\vect{V}^H \vect{V}) + \trace(\vect{Z}) \leq P, \\
    & \quad\quad \norm{\vect{U}}_F^2 \leq 1,\, 1 \leq N_d \leq \min\{M,N\},\\
    & \quad\quad \lvert \phi_{\ell} \rvert = 1 \,\, \forall \ell = 1,2,\dots,L, \\
\end{split}
\end{align*}
where $P$ denotes the total transmit power budget. This non-convex problem is solved via the following AO approach. We perform an exhaustive search over all $\min\{M,N\}$ possible values for $N_d$ in order to find the one maximizing $\mathcal{OP}_{\rm L}$'s objective. In particular, for each feasible $N_d$ value, we transform $\mathcal{OP}_{\rm L}$'s objective function into the more tractable form of the following Lemma \cite{Shi_2011_all}, and perform AO over the involved variables, as described in the sequel.
\begin{Lem} \label{Lemma_AO_AN}
Suppose that $\vect{M} \in \mathbb{C}^{N \times N}$ with $\vect{M}\succeq\vect{0}$ is:
\begin{equation} \label{eq:MSE_matrix}
\vect{M} \triangleq \left(\vect{A}^H \vect{B}\vect{C} - \vect{I}_N\right) \left(\vect{A}^H \vect{B}\vect{C} - \vect{I}_N\right)^H + \vect{A}^H\vect{R}\vect{A},
\end{equation}
where $\vect{A} \in \mathbb{C}^{M \times N}$, $\vect{B} \in \mathbb{C}^{M \times N}$, $\vect{C} \in \mathbb{C}^{N \times N}$, and $\vect{R} \in \mathbb{C}^{M \times M}$ with $\vect{R}\succ\vect{0}$. Let also the scalar function $f(\vect{S},\vect{A})\triangleq \log\lvert\vect{S}\rvert - \trace(\vect{S}\vect{M}) + \trace(\vect{I}_N)$ with $\vect{S} \in \mathbb{C}^{N \times N}$. The following maximum values for $f(\vect{S},\vect{A})$ hold: 
\begin{align}
 &\log \lvert \vect{M}^{-1} \rvert = \max_{\vect{S} \succ \vect
 {0}} f(\vect{S},\vect{A}), \label{eq:lemma1-first-part} \\
& \log \left\lvert \vect{I}_N + (\vect{B} \vect{C})^H \vect{R}^{-1} \vect{B} \vect{C} \right\rvert = \max_{\vect{A}, \vect{S} \succ \vect{0}} f(\vect{S},\vect{A}), \label{eq:lemma1-second-part}
\end{align}
where the optimal values of \eqref{eq:lemma1-first-part} and \eqref{eq:lemma1-second-part} are obtained with the solution $\vect{S}_{\rm opt} = \vect{M}^{-1}$.
\end{Lem}

By introducing the auxiliary matrix variables $\vect{A}_1,\, \vect{S}_1 \in \mathbb{C}^{N_d \times N_d}$, and defining the Mean Squared Error (MSE) matrix:
\begin{equation} \label{eq:MSE_matrix_RX}
\begin{aligned}
 \vect{M}_1 \triangleq &\left(\vect{A}_1^H \vect{U}^H \tilde{\vect{H}} \vect{V} \!-\! \vect{I}_{N_d} \right) \left(\vect{A}_1^H \vect{U}^H \tilde{\vect{H}} \vect{V} \!-\! \vect{I}_{N_d} \right)^H \\
 &+ \vect{A}_1^H \left( \vect{U}^H \left(\sigma^2 \vect{I}_M + \tilde{\vect{H}} \vect{Z} \tilde{\vect{H}}^H\right) \vect{U} \right) \vect{A}_1,
\end{aligned}
\end{equation}
$\mathcal{R}_{\rm RX}$ can be equivalently rewritten as
\begin{equation} \label{eq:equiv_R_RX}
 \mathcal{R}_{\rm RX} = \max_{\vect{A}_1, \vect{S}_1 \succ \vect{0}} \log\lvert \vect{S}_1 \rvert - \trace(\vect{S}_1 \vect{M}_1) + N_d.
\end{equation}
To apply Lemma \ref{Lemma_AO_AN} for $\hat{\mathcal{R}}_{\rm E}$, we first simplify it by noting after some manipulations that $\hat{\mathcal{R}}_{\rm E} = - \hat{\mathcal{R}}_{\rm E,1} + \hat{\mathcal{R}}_{\rm E,2}$, where
\begin{align}
 \hat{\mathcal{R}}_{\rm E,1} &= \log \left\lvert \vect{I}_K + \sigma^{-2} \vect{H}_{\rm E} \vect{Z}\vect{H}_{\rm E}^H \right\rvert, \label{eq:R_E_1}\\
 \hat{\mathcal{R}}_{\rm E,2} &= \log \left\lvert \vect{I}_K \!+\! \sigma^{-2}  \vect{H}_{\rm E} \vect{V} \vect{V}^H \vect{H}_{\rm E}^H \!+\! \sigma^{-2} \vect{H}_{\rm E} \vect{Z} \vect{H}_{\rm E}^H \right\rvert. \label{eq:R_E_2}
\end{align}
Then, by defining $\vect{Z} \triangleq \tilde{\vect{Z}} \tilde{\vect{Z}}^H$ and introducing the auxiliary variables $\vect{A}_2 \in \mathbb{C}^{K \times N}, \vect{S}_2 \in \mathbb{C}^{N \times N}$ and $\vect{S}_3 \in \mathbb{C}^{K \times K}$, \eqref{eq:R_E_1} can be re-expressed as
\begin{equation} \label{eq:R_E_part1}
 \hat{\mathcal{R}}_{{\rm E},1} = \max_{\vect{A}_2, \vect{S}_2 \succ \vect{0}} \log \lvert\vect{S}_2 \rvert - \trace\left( \vect{S}_2 \vect{M}_2 \right) + N ,
\end{equation}
with $\vect{M}_2$ being the following MSE matrix:
\begin{equation}
 \vect{M}_2 \triangleq ( \vect{A}_2^H \vect{H}_{\rm E} \tilde{\vect{Z}} - \vect{I}_N )( \vect{A}_2^H \vect{H}_{\rm E} \tilde{\vect{Z}} - \vect{I}_N)^H + \sigma^2 \vect{A}_2^H \vect{A}_2.
\end{equation}
Similarly, \eqref{eq:R_E_2} can be re-expressed as the optimization:
\begin{equation} \label{eq:R_E_part2}
 -\hat{\mathcal{R}}_{{\rm E},2} = \max_{\vect{S}_3 \succ \vect{0}} \log \lvert \vect{S}_3 \rvert -\trace\left( \vect{S}_3 \vect{M}_3 \right) +K,
\end{equation}
where $\vect{M}_3 = \vect{I}_K + \sigma^{-2} \vect{H}_{\rm E} \vect{V} \vect{V}^H \vect{H}_{\rm E}^H + \sigma^{-2}\vect{H}_{\rm E}  \tilde{\vect{Z}} \tilde{\vect{Z}}^H \vect{H}_{\rm E}^H$. The $\mathcal{OP}_{\rm L}$, excluding the optimization over $N_d$, is thus recast 
\begin{align*}
\begin{split}
\mathcal{OP}_{{\rm L},\mathbb{X}}: \,\, \max_{\mathbb{X}} \quad & \bar{\mathcal{R}}_{\rm s}\triangleq\mathcal{R}_{\rm RX} + \hat{\mathcal{R}}_{{\rm E},1} - \hat{\mathcal{R}}_{{\rm E},2} \\
\text{s.t.} \quad & \trace(\vect{V} \vect{V}^H) + \trace(\tilde{\vect{Z}}\tilde{\vect{Z}}^H) \leq P, \\
\quad & \norm{\vect{U}}_F^2 \leq 1 \\
\quad & \lvert \phi_\ell \rvert = 1 \,\, \forall \ell = 1,2,\dots,L,
\end{split}
\end{align*}
where $\mathbb{X} \triangleq \{ \vect{A}_i, \vect{S}_j \succ \vect{0}, \vect{U}, \vect{V}, \tilde{\vect{Z}}, \vect{\phi} \}$, with $i \in \{1,2\}$ and $j \in \{1,2,3\}$. $\mathcal{OP}_{{\rm L},\mathbb{X}}$ is still non-convex, due to the coupled variables, as well as the unit-modulus constraints. However, it is easy to see that it is convex when treating each set of variables separately (except $\vect{\phi}$), by a block coordinate descent approach, as presented in the following.

\subsection{$\mathcal{OP}_{\rm L}$'s Optimization with Respect to $\{\vect{A}_i\}$} \label{Sec:Optimize_A_i}
After some algebraic manipulations with \eqref{eq:equiv_R_RX} and \eqref{eq:R_E_part1} and then setting their first order derivatives with respect to $\vect{A}_1$ and $\vect{A}_2$, respectively, equal to zero, their optimal values become 
\begin{equation} \label{eq:optimal_A_1}
 \begin{aligned} 
 \vect{A}_{1,\rm opt} = \Bigl( \vect{U}^H \bigl(\sigma^2 \vect{I}_M &+  \tilde{\vect{H}} \vect{V} \vect{V}^H \tilde{\vect{H}}^H \\
 &+ \tilde{\vect{H}}\tilde{\vect{Z}} \tilde{\vect{Z}}^H \tilde{\vect{H}}^H \bigr) \vect{U} \Bigr)^{-1} \vect{U}^H \tilde{\vect{H}} \vect{V},
\end{aligned}
\end{equation} 
\begin{equation} \label{eq:optimal_A_2}
  \vect{A}_{2, \rm opt} = \left( \sigma^2 \vect{I}_K + \vect{H}_{\rm E} \tilde{\vect{Z}} \tilde{\vect{Z}}^H \vect{H}_{\rm E}^H \right)^{-1} \vect{H}_{\rm E} \tilde{\vect{Z}}.
\end{equation}

\subsection{$\mathcal{OP}_{\rm L}$'s Optimization with Respect to $\{\vect{S}_j\}$} \label{Sec:Optimize_S_j}
By substituting $\vect{A}_{1, \rm opt}$ and $\vect{A}_{2, \rm opt}$ into \eqref{eq:equiv_R_RX} and \eqref{eq:R_E_part1}, respectively, and invoking the matrix inversion lemma, the optimal expressions for $\vect{S}_1$ and $\vect{S}_2$ are obtained as
\begin{equation} \label{eq:optimal_S_1}
 \begin{aligned}
 \vect{S}_{1, \rm opt} = \vect{I}_{N_d} &+ \vect{V}^H \tilde{\vect{H}}^H \vect{U} \Bigl( \sigma^2 \vect{U}^H \vect{U} \\
 &+ \vect{U}^H \tilde{\vect{H}} \tilde{\vect{Z}} \tilde{\vect{Z}}^H \tilde{\vect{H}}^H \vect{U} \Bigr)^{-1} \vect{U}^H \tilde{\vect{H}} \vect{V} ,
 \end{aligned}
\end{equation}
\begin{equation} \label{eq:optimal_S_2}
  \vect{S}_{2, \rm opt} = \vect{I}_N + \sigma^{-2} \tilde{\vect{Z}}^H \vect{H}_{\rm E}^H \vect{H}_{\rm E} \tilde{\vect{Z}}.
\end{equation}
The optimal $\vect{S}_3$ is obtained by Lemma \ref{Lemma_AO_AN} as $\vect{S}_{3, \rm opt} = \vect{M}_3^{-1}$.

\subsection{$\mathcal{OP}_{\rm L}$'s Optimization with Respect to $\vect{U}$} \label{Sec:Optimize_U}
The optimization variable $\vect{U}$ appears only in the expression $\mathcal{R}_{\rm RX}$. It, hence, suffices to obtain its Lagrangian function and then equate its first-order derivative with zero. For $\kappa \geq 0$ being the Lagrange multiplier, the Lagrangian of $\vect{U}$ is:
\begin{equation} \label{eq:Lagrangian_U}
\mathcal{L}_{\mathcal{OP}_{L,\vect{U}}}(\vect{U},\kappa) = -\trace(\vect{S}_1 \vect{M}_1) - \kappa (\trace(\vect{U}^H \vect{U}) - 1).
\end{equation}
After replacing $\vect{M}_1$ and treating the terms irrelevant to $\vect{U}$ as constants, the linear system $\vect{E} \vect{U} \vect{F} + \kappa \vect{U} = \vect{J}$ is obtained, where
$\vect{E} \triangleq \sigma^2 \vect{I}_M + \tilde{\vect{H}} \vect{V} \vect{V}^H \tilde{\vect{H}}^H + \tilde{\vect{H}} \tilde{\vect{Z}} \tilde{\vect{Z}}^H \tilde{\vect{H}}^H$, $\vect{F} \triangleq \vect{A}_1 \vect{S}_1 \vect{A}_1^H$, and $\vect{J} \triangleq \tilde{\vect{H}} \vect{V} \vect{S}_1 \vect{A}_1^H $. The optimal $\vect{U}$ is then derived as follows:
\begin{equation} \label{eq:semi_optimal_U}
 \operatorname{vec}(\vect{U}^{\star}) = \left( \vect{F}^T \otimes \vect{E} + \kappa \vect{I}_{M N_d} \right)^{-1} \operatorname{vec}(\vect{J}),
\end{equation}
\begin{equation} \label{eq:optimal_U}
 \vect{U}_{\rm opt}^{\kappa} = \left( \operatorname{vec}(\vect{I}_{N_d})^T \otimes \vect{I}_M \right) (\vect{I}_{N_d} \otimes \operatorname{vec}(\vect{U}^{\star}) ).
\end{equation}
It can be observed from \eqref{eq:semi_optimal_U} and \eqref{eq:optimal_U} that $\vect{U}$ depends on $\kappa$. To ensure the complementary slackness condition \cite{Boyd_2004}:
\begin{equation}
 \kappa^{\star} \left( \trace\left((\vect{U}_{\rm opt}^{\kappa^{\star}})^H \vect{U}_{\rm opt}^{\kappa^{\star}}\right)  - 1 \right) = 0,
\end{equation}
the optimal $\kappa$, denoted by $\kappa^{\star}$, can be computed using the following Corollary.

\begin{Cor} \label{Thm:OP_L_Find_kappa}
Let $\vect{Q} \vect{\Xi} \vect{Q}^H$ be the eigendecomposition of $\vect{F}^T \otimes \vect{E}$, i.e., $\vect{\Xi}$ is a $M N_d \times M N_d$ diagonal matrix whose elements are the eigenvalues of $\vect{F}^T \otimes \vect{E}$ and $\vect{Q} \in \mathbb{C}^{M N_d \times M N_d}$ contains the corresponding eigenvectors. The Lagrange multiplier $\kappa^{\star}$ can be obtained from the solution of the equation:
\begin{equation} \label{eq:optimum_kappa}
 \sum_{p = 1}^{M N_d} \frac{[\tilde{\vect{Q}}]_{p,p}}{([\vect{\Xi}]_{p,p} + \kappa)^2} = N_d^{-1},
\end{equation}
where $\tilde{\vect{Q}} \triangleq \vect{Q}^H \operatorname{vec}(\vect{J}) \operatorname{vec}(\vect{J})^H \vect{Q}$.
\end{Cor}

\begin{IEEEproof}
Omitted due to space limitations.
\end{IEEEproof}
It can be easily observed that \eqref{eq:optimum_kappa}'s left-hand side is monotonically decreasing for $\kappa \geq 0$. Hence, $\kappa^{\star}$ can be obtained using a one-dimensional search, e.g., the bisection method. Once $\kappa^{\star}$ is computed, it can be replaced in \eqref{eq:semi_optimal_U} to get the optimal $\vect{U}_{\rm opt}$, as shown in \eqref{eq:optimal_U}.

\subsection{$\mathcal{OP}_{\rm L}$'s Optimization with Respect to $\{\vect{V}, \tilde{\vect{Z}} \}$} \label{Sec:Optimize_V_Z}
For the optimization over $\vect{V}$ and $\tilde{\vect{Z}}$, it suffices to use the Lagrangian function of the reformulated objective $\bar{\mathcal{R}}_s$ in $\mathcal{OP}_{\rm L, \mathbb{X}}$ and set its first-order derivatives with respect to $\vect{V}$ and $\tilde{\vect{Z}}$, respectively, equal to zero, resulting in the expressions:
\begin{align}
 \vect{V}_{\rm opt}^{\lambda} &= (\lambda \vect{I}_N + \vect{R}_{\vect{V}_1} )^{-1} \vect{R}_{\vect{V}_2} \label{eq:OP_L_V_opt} \\
  \tilde{\vect{Z}}_{\rm opt}^{\lambda} &= (\lambda \vect{I}_N + \vect{R}_{\tilde{\vect{Z}}_1} )^{-1} \vect{R}_{\tilde{\vect{Z}}_2}, \label{eq:OP_L_Z_opt}
\end{align}
where $\mathbf{K}\triangleq\tilde{\vect{H}}^H \vect{U} \vect{A}_1 \vect{S}_1 \vect{A}_1^H \vect{U}^H \tilde{\vect{H}}$, $\vect{R}_{\tilde{\vect{Z}}_2} \triangleq \vect{H}_{\rm E}^H \vect{A}_2 \vect{S}_2$, and
\begin{align}
 \vect{R}_{\vect{V}_1} &\triangleq \mathbf{K} + \sigma^{-2} \vect{H}_{\rm E}^H \vect{S}_3 \vect{H}_{\rm E}, \\
 \vect{R}_{\vect{V}_2} &\triangleq \tilde{\vect{H}}^H \vect{U} \vect{A}_1 \vect{S}_1, \\
\vect{R}_{\tilde{\vect{Z}}_1} &\triangleq \mathbf{K}+\vect{H}_{\rm E}^H \vect{A}_2 \vect{S}_2 \vect{A}_2^H \vect{H}_{\rm E}+ \sigma^{-2} \vect{H}_{\rm E}^H \vect{S}_3 \vect{H}_{\rm E}.
\end{align}
Clearly, the optimum pair $(\vect{V}_{\rm opt}^{\lambda}, \tilde{\vect{Z}}_{\rm opt}^{\lambda})$ depends on $\lambda$. Similar to the optimization with respect to $\vect{U}$ using Corollary~\ref{Thm:OP_L_Find_kappa}, by defining the eigendecompositions $\vect{R}_{\vect{V}_1}\triangleq\vect{P}_{\vect{V}} \vect{\Lambda}_{\vect{V}} \vect{P}_{\vect{V}}^H$ and $\vect{R}_{\tilde{\vect{Z}}_1} \triangleq \vect{P}_{\tilde{\vect{Z}}} \vect{\Lambda}_{\tilde{\vect{Z}}} \vect{P}_{\tilde{\vect{Z}}}^H$, the optimum $\lambda$ can be obtained by solving the following equation via a bisection algorithm:
\begin{equation}
 \sum_{q = 1}^N \left( \frac{[\tilde{\vect{P}}_{\vect{V}}]_{q,q}}{\left( [\vect{\Lambda}_{\vect{V}}]_{q,q} + \lambda \right)^2} + \frac{[\tilde{\vect{P}}_{\tilde{\vect{Z}}}]_{q,q}}{\left( [\vect{\Lambda}_{\tilde{\vect{Z}}}]_{q,q} + \lambda \right)^2} \right) = P,
\end{equation}
where $\tilde{\vect{P}}_{\vect{V}} \triangleq \vect{P}_{\vect{V}}^H \vect{R}_{\vect{V}_2} \vect{R}_{\vect{V}_2}^H \vect{P}_{\vect{V}}$ and $\tilde{\vect{P}}_{\tilde{\vect{Z}}} \triangleq \vect{P}_{\tilde{\vect{Z}}}^H \vect{R}_{\tilde{\vect{Z}}_2} \vect{R}_{\tilde{\vect{Z}}_2}^H \vect{P}_{\tilde{\vect{Z}}}$.

\begin{algorithm}[!t]
\begin{algorithmic}[1]
\caption{Proposed Solution for $\mathcal{OP}_{{\rm L},\vect{\phi}}$}
\label{alg:OP_L_phi}
\State \textbf{Input:} $\vect{A}_i$ with $i\in\{1,2\}$, $\vect{S}_j$ with $j\in\{1,2,3\}$, $\vect{U}$, $\vect{V}$, $\vect{Z}$, $\epsilon>0$, $\kappa > 0$, $\mu,\nu\in (0,1)$, and $\vect{\phi}_0$.
\State Compute $\vect{q}_0 = - \nabla_{\vect{\phi}_0}^{\rm R} g$.
\For{$ n = 1, 2, \dots$}
    \State The Armijo-Goldstein backtracking line search:
    
    \hspace{-0.29cm} Find the smallest integer $\omega \geq 0$ such that
    
    \hspace{-0.29cm}\vspace*{-\baselineskip}
          \begin{fleqn}[\dimexpr(\leftmargini-\labelsep)*2]
            \setlength\belowdisplayskip{0pt}
            \begin{equation*}
                \begin{multlined}[c]
                g\left(\operatorname{unit}(\vect{\phi}_{n-1} + \kappa\nu^{\omega}\vect{q}_{n-1}) \right) - g(\vect{\phi}_{n-1}) \\
                \leq \mu \kappa \nu^{\omega} \Re\{ (\nabla_{\vect{\phi}_{n-1}}^{\rm R} g)^H \vect{q}_{n-1} \}.
                \end{multlined}
            \end{equation*}
          \end{fleqn}    
    \State Compute $\tau_{n-1} = \kappa \nu^{\omega}$.    
    \State Compute $\tilde{\vect{\phi}}_n = \vect{\phi}_{n-1} + \tau_{n-1} \vect{q}_{n-1}$   
    
    \hspace{-0.29cm} and $\vect{\phi}_n = \operatorname{unit}(\tilde{\vect{\phi}}_n)$.    
    \State Compute $\vect{q}_n$ and the Polak-Ribière constant $\zeta_{n-1}$
    
    \hspace{-0.29cm} according to \eqref{eq:CG_direction_phi} and \eqref{eq:Polak_Ribiere_zeta_n}, respectively.
    
    \If $\norm{\nabla_{\vect{\phi}_n}^{\rm R}}^2 \leq \epsilon$
        \State $\vect{\phi}^{\star} = \vect{\phi}_n$ and \textbf{break};   
    \EndIf
\EndFor
\State \textbf{Output:} $\vect{\phi}^{\star}$.
\end{algorithmic}
\end{algorithm}

\subsection{$\mathcal{OP}_{\rm L}$'s Optimization with Respect to $\vect{\phi}$} \label{Sec:Optimize_phi}
The optimization variable $\vect{\phi}$ in $\mathcal{OP}_{\rm L}$ appears only in the legitimate rate expression $\bar{\mathcal{R}}_{\rm s}$. By keeping only the terms that depend on $\vect{\phi}$ to obtain $g\triangleq-\bar{\mathcal{R}}_{\rm s}(\vect{\phi})$, $\mathcal{OP}_{\rm L}$'s optimization over this variable reduces to:
\begin{align*}
\mathcal{OP}_{{\rm L},\vect{\phi}}: \,\,\min_{\vect{\phi}} \quad  g \quad \text{s.t.} \quad \lvert \phi_\ell \rvert = 1 \,\, \forall \ell = 1,2,\dots,L.
\end{align*}
We adopt Riemannian-Manifold optimization \cite{Absil_2008} to solve this problem via Algorithm~\ref{alg:OP_L_phi}; detailed derivations will be provided in the extended version of this paper. At each $n$-th iterative step of this algorithm, $\vect{q}_n$ is derived as
\begin{equation} \label{eq:CG_direction_phi}
 \vect{q}_n = - \nabla_{\vect{\phi}_n}^{\rm R} g + \zeta_{n-1} \mathcal{T}_{n-1 \rightarrow n}(\vect{q}_{n-1}),
\end{equation}
where $\vect{\phi}_n$ represents the legitimate RIS reflection vector at the $n$-th step, and $\mathcal{T}_{n-1 \rightarrow n}$ is defined for any vector $\vect{r}$ as
\begin{equation}
 \mathcal{T}_{n-1 \rightarrow n}(\vect{r}) \triangleq \vect{r} - \Re \{ \vect{r} \odot (\vect{\phi}_n^T)^H \} \odot \vect{\phi}_n.
\end{equation}
In addition, $\zeta_{n-1}$ is the Polak-Ribière parameter given by 
\begin{equation} \label{eq:Polak_Ribiere_zeta_n}
\! \zeta_{n-1} \!=\! \frac{\Re{\left\{\left( \nabla_{\vect{\phi}_n}^{\rm R} g \right)^H \left( \nabla_{\vect{\phi}_n}^{\rm R} g - \mathcal{T}_{n-1 \rightarrow n}(\nabla_{\vect{\phi}_{n-1}}^{\rm R} g) \right)  \right\}}}{\norm{\nabla_{\vect{\phi}_{n-1}}^{\rm R} g}^2}.
\end{equation}

All steps solving $\mathcal{OP}_{\rm L}$, using the AO described in the previous subsections, are summarized in Algorithm \ref{alg:OP_L_Overall_Algorithm}, whose convergence proof is omitted due to space limitations.

\begin{algorithm}[!t]
\begin{algorithmic}[1]
\caption{Proposed Secrecy Design Solving $\mathcal{OP}_{\rm L}$}
\label{alg:OP_L_Overall_Algorithm}
\State \textbf{Input:} $p=0$, $\epsilon > 0$, as well as feasible $\vect{V}^{(0)}$, $\vect{Z}^{(0)}$, $\vect{\phi}^{(0)}$, and $\hat{\mathcal{R}}_{\rm s}^{(0)}$ as defined in $\mathcal{OP}_{\rm L}$.
\For{$ m = 1,2,\ldots,\min\{M,N\}$} 
    \For{$ p = 1,2,\dots$}
    \State Compute $\tilde{\vect{H}} = \vect{H}_2 \diag{\{\vect{\phi}^{(p-1)}\}} \vect{H}_1$.
    \State Compute $\vect{A}_i$ with $i \in \{1,2\}$ using \eqref{eq:optimal_A_1}, \eqref{eq:optimal_A_2}.
		\State Compute $\vect{S}_{1}$ using \eqref{eq:optimal_S_1}, $\vect{S}_{2}$ using \eqref{eq:optimal_S_2}, 
		
		\hspace{0.35cm}and $\vect{S}_3 = \vect{M}_3^{-1}$.
    \State Compute $\vect{U}^{(p)}_m$ using \eqref{eq:semi_optimal_U}, \eqref{eq:optimal_U}, and 
    
    \hspace{0.35cm}a bisection method.
    \State Compute $\vect{V}^{(p)}_m$ and $\tilde{\vect{Z}}^{(p)}$ according to \eqref{eq:OP_L_V_opt}, \eqref{eq:OP_L_Z_opt}, 
    
    \hspace{0.25cm} and a bisection method.
		\State Set $\vect{Z}^{(p)}_m = \tilde{\vect{Z}}^{(p)} \left(\tilde{\vect{Z}}^{{(p)}}\right)^H$.
    \State Obtain $\vect{\phi}^{(p)}_m$ using Algorithm \ref{alg:OP_L_phi}.
    \If $\left\lvert\left(\hat{\mathcal{R}}_{\rm s}^{(p)} - \hat{\mathcal{R}}_{\rm s}^{(p-1)}\right)/\hat{\mathcal{R}}_{\rm s}^{(p)}\right\rvert \leq \epsilon$, \textbf{break}; 
    \EndIf
    \EndFor
\State Compute $\hat{\mathcal{R}}_{\rm s}^{(p)}$ for $N_d=m$ streams using 

\hspace{-0.19cm}$\vect{U}_m^{(p)}$, $\vect{V}_m^{(p)}$, $\vect{Z}_m^{(p)}$, and $\vect{\phi}_m^{(p)}$.
\EndFor
\State Choose $N_d = m^{\star}$ with $m^{\star}$ yielding the maximum rate.
\State \textbf{Output:} $\vect{U}_{m^{\star}}^{(p)}$, $\vect{V}_{m^{\star}}^{(p)}$, $\vect{Z}_{m^{\star}}^{(p)}$, and $\vect{\phi}_{m^{\star}}^{(p)}$.
\end{algorithmic}
\end{algorithm}

\section{Numerical Results and Discussion} \label{Sec:Numerical} 
In this section, we investigate the secrecy rate performance of the proposed PLS scheme over frequency flat Rayleigh fading channels with zero mean and unit variance, and for distance-dependent pathloss with exponent equal to $2$ for all involved links. We have particularly evaluated the achievable rates of the legitimate and eavesdropping links using expressions \eqref{eq:R_RX} for $\mathcal{R}_{\rm RX}$ and \eqref{eq:R_E} for $\mathcal{R}_{\rm E}$, respectively, and $\mathcal{R}_{\rm s}$ providing the achievable secrecy rate. We have adopted the proposed PLS scheme in Section~\ref{E_design} for the receive combining and the RIS passive beamforming of the eavesdropping system. For the legitimate system, we have used the proposed PLS scheme in Section~\ref{Sec:Prob_Form} encompassing BS precoding and AN, receive combining, and legitimate RIS passive beamforming, as well as a special version of it for the case where a legitimate RIS is not available. For this special version, we have solved a similar problem to $\mathcal{OP}_{\rm L}$ via Lemma \ref{Lemma_AO_AN} and AO, by removing the links involving the legitimate RIS and the optimization over its relevant variable $\vect{\phi}$. In our simulations, the BS was located in the origin of the $xy$ plane, whereas RX and E lied on a circle of radius $10m$ in the angles $45^{\rm o}$ and $85^{\rm o}$, respectively, from BS. The first unit element of the eavesdropping RIS was placed in the middle of the line connecting RX and E, and the other elements expand along the positive directions of the $x$ and $y$ axes. In a similar manner, the legitimate RIS is placed in the same circle as RX and E, in the angle $20^{\rm o}$ from BS. In addition, we have used the following parameters' setting: $N =\{8,16\}$, $M=K=4$, $\sigma^2 = 1$, and $200$ independent Monte Carlo realizations. 
For comparison purposes, we have implemented the baseline scheme proposed in \cite{Hong_2020} that optimizes the BS precoding matrix and AN for fixed numbers $N_d$ of the data streams, and assumes optimum RX combining.  

\begin{figure}[!t]
\centering
\includegraphics[scale=0.58]{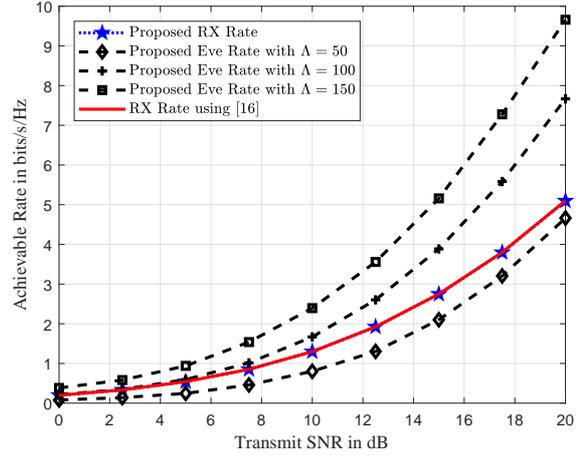}
\caption{Achievable rates in bps/Hz at the legitimate RX and the eavesdropper E versus the transmit SNR in dB for $N=8$ BS antennas and different numbers $\Lambda$ for the unit elements at the eavesdropping RIS. The legitimate system does not possess a RIS trying to safeguard confidential communication with only BS precoding and AN, and RX combining.}
\label{fig:AN_only}
\end{figure}

We commence in Fig$.$~\ref{fig:AN_only} with the achievable rate performance in bps/Hz for the legitimate and eavesdropping links as functions of the BS transmit SNR, defined as $P/\sigma^2$, using $N=8$ BS antennas. For these results, we have considered that the legitimate system does not include a RIS, and targets at securing confidential transmissions with only BS precoding and AN. It can be seen that the rates increase with increasing SNR for both the legitimate and eavesdropping links. It is depicted that RX's rate is larger than E's rate for $\Lambda=50$, however, when the larger simulated values for $\Lambda$ are considered, E's rate is similar or larger than that of RX. This reveals that, with the proposed schemes for RX and E, the secrecy rate equals to $0$ for cases of existence of an eavesdropping RIS with $\Lambda>50$ unit elements. For such cases, BS precoding and AN are incapable of safeguarding the legitimate link. This holds for both the proposed design and the baseline scheme, whose performances coincide. Recall that \cite{Hong_2020} considers optimum RX combining, which is non linear, while our scheme provides a linear combiner. The latter behavior happens due to the fact that BS is unaware of the presence of the eavesdropping RIS (which can have $\Lambda\gg N$ unit elements \cite{HMIMO_all}), and only possesses $\mathbf{H}_{\rm E}$ for the design of the legitimate link's parameters. 

\begin{figure}[!t]
\centering
\includegraphics[scale=0.58]{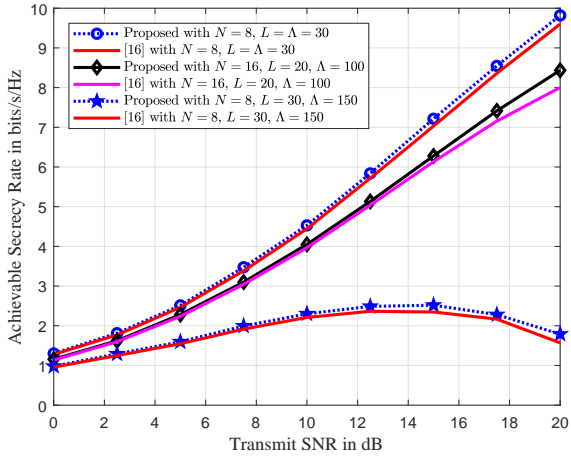}
\caption{Achievable rates in bps/Hz versus the transmit SNR in dB for $N=\{8,16\}$ antenna elements at the legitimate BS, different numbers $L$ of the unit elements at the legitimate RIS, and different numbers $\Lambda$ for the unit elements of the eavesdropping RIS. In contrast to Fig$.$\ref{fig:AN_only}, the legitimate system safeguards communication with BS precoding and AN, RX combining, and RIS passive beamforming.}
\label{fig:RIS_RX}
\end{figure}

In Fig$.$~\ref{fig:RIS_RX}, we consider that the legitimate system deploys a RIS with $L$ unit elements and applies the joint design of Algorithm~2 in Section~\ref{Sec:Prob_Form}. We have plotted the achievable secrecy rates in bps/Hz versus the transmit SNR in dB for $N=\{8,16\}$ BS antennas and different numbers $L$ and $\Lambda$ for the unit elements of the legitimate and eavesdropping RISs, respectively. As shown using the proposed design and the baseline scheme \cite{Hong_2020} for $N_d=\min\{8,4\}=4$, the consideration of a legitimate RIS, with even more than $500\%$ less elements than the eavesdropping one, results in positive secrecy rates for all considered SNR values. It can be observed that, when $\Lambda=5L$, there exists a moderate SNR value where the secrecy rate gets its maximum value. For the example with $N=8$, $L=30$, and $\Lambda=150$, the maximum secrecy rate is $2.5$bits/s/Hz at the SNR value $15$dB. Below this value, the rate increases with increasing SNR, while above this value, an SNR increase results in smaller rate. This happens because the eavesdropping capability offered by the $5L$-element eavesdropping RIS increases, while it cannot be treated by the $L$-element legitimate RIS. Interestingly, it is also depicted that the proposed scheme, that utilizes linear RX combining and the optimum $N_d$ value for each considered SNR value, outperforms \cite{Hong_2020} with $N_d=4$.

\section{Conclusion} \label{Sec:Conclusion}
In this paper, we studied RIS-empowered MIMO PLS communications, where RISs are deployed from both the legitimate and the eavesdropping systems. We focused on the case where the RISs are placed close to the receivers and their existence is unknown to the competing system. A joint design of legitimate precoding with AN, RX combining, and passive legitimate RIS beamforming was presented that was shown capable of safeguarding MIMO communication over RIS-empowered eavesdropping systems, with an around $500\%$ larger RIS compared to the legitimate one. 

\bibliographystyle{IEEEtran}
\bibliography{references}

\begin{thebibliography}{10}
\providecommand{\url}[1]{#1}
\csname url@samestyle\endcsname
\providecommand{\newblock}{\relax}
\providecommand{\bibinfo}[2]{#2}
\providecommand{\BIBentrySTDinterwordspacing}{\spaceskip=0pt\relax}
\providecommand{\BIBentryALTinterwordstretchfactor}{4}
\providecommand{\BIBentryALTinterwordspacing}{\spaceskip=\fontdimen2\font plus
\BIBentryALTinterwordstretchfactor\fontdimen3\font minus
  \fontdimen4\font\relax}
\providecommand{\BIBforeignlanguage}[2]{{%
\expandafter\ifx\csname l@#1\endcsname\relax
\typeout{** WARNING: IEEEtran.bst: No hyphenation pattern has been}%
\typeout{** loaded for the language `#1'. Using the pattern for}%
\typeout{** the default language instead.}%
\else
\language=\csname l@#1\endcsname
\fi
#2}}
\providecommand{\BIBdecl}{\relax}
\BIBdecl

\bibitem{Liaskos_Visionary_2018_all}
C.~Liaskos, S.~Nie, A.~I. Tsioliaridou, A.~Pitsillides, S.~Ioannidis, and I.~F.
  Akyildiz, ``{A new wireless communication paradigm through
  software-controlled metasurfaces},'' \emph{IEEE Commun. Mag.}, vol.~56,
  no.~9, pp. 162--169, Sep. 2018.

\bibitem{George_RIS_TWC2019_all}
C.~Huang, A.~Zappone, G.~C. Alexandropoulos, M.~Debbah, and C.~Yuen,
  ``{Reconfigurable intelligent surfaces for energy efficiency in wireless
  communication},'' \emph{IEEE Trans. Wireless Commun.}, vol.~18, no.~8, pp.
  4157--4170, Aug. 2019.

\bibitem{Wu_RIS_TWC2019}
Q.~Wu and R.~Zhang, ``{Intelligent reflecting surface enhanced wireless network
  via joint active and passive beamforming},'' \emph{IEEE Trans. Wireless
  Commun.}, vol.~18, no.~11, pp. 5394--5409, Nov. 2019.

\bibitem{Kaina_metasurfaces_2014_all}
N.~Kaina, M.~Dupre, G.~Lerosey, and M.~Fink, ``Shaping complex microwave fields
  in reverberating media with binary tunable metasurfaces,'' \emph{Sci. Rep.
  4}, pp. 1--7, Article No 076401, 2014.

\bibitem{Marco_Visionary_2019_all1}
M.~Di~Renzo, M.~Debbah, D.-T. Phan-Huy, A.~Zappone, M.-S. Alouini, C.~Yuen,
  V.~Sciancalepore, G.~C. Alexandropoulos, J.~Hoydis, H.~Gacanin, J.~de~Rosny,
  A.~Bounceu, G.~Lerosey, and M.~Fink, ``{Smart radio environments empowered by
  AI reconfigurable meta-surfaces: An idea whose time has come},''
  \emph{EURASIP J. Wireless Commun. Netw.}, vol. 129, pp. 1--20, May 2019.

\bibitem{HMIMO_all}
C.~Huang, S.~Hu, G.~C. Alexandropoulos, A.~Zappone, C.~Yuen, R.~Zhang,
  M.~Di~Renzo, and M.~Debbah, ``Holographic {MIMO} surfaces for {6G} wireless
  networks: {O}pportunities, challenges, and trends,'' \emph{IEEE Wireless
  Commun.}, to appear, 2020.

\bibitem{Yang_ComMag_2015_all}
N.~Yang, L.~Wang, G.~Geraci, M.~Elkashlan, J.~Yuan, and M.~Di~Renzo,
  ``Safeguarding {5G} wireless communication networks using physical layer
  security,'' \emph{IEEE Commun. Mag.}, vol.~53, no.~4, pp. 20--27, Apr. 2015.

\bibitem{Chen_2019_all}
J.~Chen, Y.-C. Liang, Y.~Pei, and H.~Guo, ``Intelligent reflecting surface: {A}
  programmable wireless environment for physical layer security,'' \emph{IEEE
  Access}, vol.~7, pp. 82\,599--82\,612, Jul. 2019.

\bibitem{Shen_2019_all}
H.~Shen, W.~Xu, S.~Gong, Z.~He, and C.~Zhao, ``\BIBforeignlanguage{en}{Secrecy
  rate maximization for intelligent reflecting surface assisted multi-antenna
  communications},'' \emph{\BIBforeignlanguage{en}{IEEE Commun. Lett.}},
  vol.~23, no.~9, pp. 1488--1492, Sep. 2019.

\bibitem{Cui_2019_all}
M.~Cui, G.~Zhang, and R.~Zhang, ``\BIBforeignlanguage{en}{Secure wireless
  communication via intelligent reflecting surface},''
  \emph{\BIBforeignlanguage{en}{IEEE Wireless Commun. Lett.}}, vol.~8, no.~5,
  pp. 1410--1414, Oct. 2019.

\bibitem{Xu_2019_all}
D.~Xu, X.~Yu, Y.~Sun, D.~W.~K. Ng, and R.~Schober, ``Resource allocation for
  secure {IRS}-assisted multiuser {MISO} systems,'' in \emph{Proc. IEEE
  GLOBECOM}, Waikoloa, USA, Dec. 2019, pp. 1--6.

\bibitem{Yu_2019_all}
X.~Yu, D.~Xu, and R.~Schober, ``Enabling secure wireless communications via
  intelligent reflecting surfaces,'' in \emph{Proc. IEEE GLOBECOM}, Waikoloa,
  USA, Dec. 2019, pp. 1--6.

\bibitem{Almohamad_2020}
A.~{Almohamad}, A.~M. {Tahir}, A.~{Al-Kababji}, H.~M. {Furqan}, T.~{Khattab},
  M.~O. {Hasna}, and H.~{Arslan}, ``Smart and secure wireless communications
  via reflecting intelligent surfaces: A short survey,'' \emph{IEEE Open J.
  Commun. Soc.}, vol.~1, pp. 1442--1456, Sep. 2020.

\bibitem{Chu_2020_all}
Z.~Chu, W.~Hao, P.~Xiao, and J.~Shi, ``\BIBforeignlanguage{en}{Intelligent
  reflecting surface aided multi-antenna secure transmission},''
  \emph{\BIBforeignlanguage{en}{IEEE Wireless Commun. Lett.}}, vol.~9, no.~1,
  pp. 108--112, Jan. 2020.

\bibitem{Lyu_2020_all}
B.~{Lyu}, D.~T. {Hoang}, S.~{Gong}, D.~{Niyato}, and D.~I. {Kim}, ``Irs-based
  wireless jamming attacks: When jammers can attack without power,'' \emph{IEEE
  Wireless Commun. Lett.}, vol.~9, no.~10, pp. 1663--1667, Oct. 2020.

\bibitem{Hong_2020}
S.~Hong, C.~Pan, H.~Ren, K.~Wang, and A.~Nallanathan, ``Artificial-noise-aided
  secure {MIMO} wireless communications via intelligent reflecting surface,''
  \emph{IEEE Trans. Commun.}, to appear, 2020.

\bibitem{Cumanan_2017_all}
K.~Cumanan, G.~C. Alexandropoulos, Z.~Ding, and G.~K. Karagiannidis, ``Secure
  communications with cooperative jamming: {O}ptimal power allocation and
  secrecy outage analysis,'' \emph{IEEE Trans. Veh. Technol.}, vol.~66, no.~8,
  pp. 7495--7505, Aug. 2017.

\bibitem{George_RIS_2020}
G.~C. Alexandropoulos and E.~Vlachos, ``A hardware architecture for
  reconfigurable intelligent surfaces with minimal active elements for explicit
  channel estimation,'' in \emph{Proc. IEEE ICASSP}, Barcelona, Spain, May
  2020, pp. 9175--9179.

\bibitem{Deepak_ICASSP_2019}
D.~Mishra and H.~Johansson, ``Channel estimation and low-complexity beamforming
  design for passive intelligent surface assisted {MISO} wireless energy
  transfer,'' in \emph{Proc. IEEE ICASSP}, Brighton, UK, May 2019, pp.
  4659--4663.

\bibitem{Liu_2017_all}
Y.~Liu, L.~Li, G.~C. Alexandropoulos, and M.~Pesavento, ``Securing relay
  networks with artificial noise: {A}n error performance based approach,''
  \emph{MDPI Entropy 19}, no. 8: 384, pp. 7495--7505, Jul. 2017.

\bibitem{Shi_2011_all}
Q.~Shi, M.~Razaviyayn, Z.-Q. Luo, and C.~He, ``\BIBforeignlanguage{en}{An
  iteratively weighted {MMSE} approach to distributed sum-utility maximization
  for a {MIMO} interfering broadcast channel},''
  \emph{\BIBforeignlanguage{en}{IEEE Trans. Signal Process.}}, vol.~59, no.~9,
  pp. 4331--4340, Sep. 2011.

\bibitem{Boyd_2004}
S.~Boyd and L.~Vandenberghe, \emph{Convex Optimization}.\hskip 1em plus 0.5em
  minus 0.4em\relax Cambridge University Press, 2004.

\bibitem{Absil_2008}
P.-A. Absil, R.~Mahony, and R.~Sepulchre, \emph{Optimization Algorithms on
  Matrix Manifolds}.\hskip 1em plus 0.5em minus 0.4em\relax Princeton
  University Press, 2008.

\end{thebibliography}

\end{document}